\documentstyle[prl,aps,epsfig,floats]{revtex}
\begin{document}

\draft

\wideabs{

\title {Electronic Transport in Metal Nanocrystal Arrays:
The Effect of Structural Disorder on Scaling Behavior}

\author{Raghuveer Parthasarathy, Xiao-Min Lin, and Heinrich M. Jaeger}
\address{James Franck Institute and Department of Physics, 
  University of Chicago,
  Chicago, IL 60637}

\date{\today}

\maketitle

\begin{abstract}
We investigate the impact of structural disorder on electronic transport
in gold nanocrystal monolayers. Arrays ranging from void-filled networks
to well-ordered superlattices show clear voltage thresholds ($V_{\rm T}$)
due to
Coulomb blockade, and temperature-independent conduction indicative of
quantum tunneling. Current-voltage characteristics of arrays with and
without long-range structural order were found to collapse onto distinct
scaling curves. The former follow a single power-law: 
$I \sim (V-V_{\rm{T}})^\zeta$, $\zeta = 2.25\pm0.1. $
The latter show additional structure, reflecting the underlying
disordered topology.
\end {abstract}


}

 	The intriguing electronic and optical properties of individual
nanocrystal quantum dots have unleashed a flood of interest \cite{timp}.
However,
despite such discoveries as metal-insulator-type transitions in squeezed
nanocrystal monolayers \cite{collier97} and 
spin-dependent transport in magnetic particle assemblies
\cite{black},
the simplest nanoparticle array --- a single layer of metal
nanocrystals --- has remained poorly understood.  The main reason is that
the transport characteristics are strongly affected by three types of
disorder:  global structural disorder in the array topology, local
structural disorder in the interparticle couplings, and local charge
disorder due to random, immobile charges in the underlying
substrate.  
Theoretical approaches investigating tunneling transport so far
have concentrated on local charge disorder only \cite{mw}.  
A full treatment of
the combined types of disorder is not available, even though large
differences between spatially ordered and disordered structures might be
expected due to the sensitivity of percolative charge transport phenomena
to array topology and the exponential dependence of local tunneling
resistances on the interparticle spacings.  With nanocrystal arrays as
``artificial solids'' \cite{collier98} 
expected to provide useful analogues and tunable
test-beds for various bulk correlated electron systems, a full
understanding of metal nanocrystal monolayers is vital.

Experimentally, a high degree of structural order has been
elusive for arrays between in-plane electrodes, and previous
investigations of electrical conduction in 2D nanocrystal systems have
been performed only on small, highly disordered, or multi-layered arrays
\cite{black,andres,cordanetal,wybourne}.  
Using newly-developed self-assembly techniques, we have fabricated
large highly-ordered monolayers of dodecanethiol-ligated gold nanocrystals
on substrates with in-plane electrodes.  Transport measurements and
subsequent transmission electron microscopy (TEM) on the same arrays
allowed for direct correlation of electronic and structural
characteristics.
Comparison of these superlattices with void-filled networks
for the first time delineates the roles played by the different
types of disorder.

Nanocrystals were deposited on silicon substrates coated with
100nm amorphous silicon nitride (Si$_3$N$_4$) (Fig. 1a).  
Under a 70$\mu$m $\times$ 70$\mu$m
area,  the Si was etched away to leave a freestanding 
Si$_3$N$_4$ membrane
``window,'' allowing TEM imaging \cite{morkved}.  
Thin (20nm) Cr electrodes were
patterned using electron-beam lithography.  1-dodecanethiol ligated gold
nanocrystals were synthesized as described in Ref. \cite{lin1} 
and dissolved in
toluene to a concentration of about $2.4 \times 10^{13}$ ml$^{-1}$.
The gold core radii
varied from sample to sample (2.2--2.9nm), but for each sample were
monodisperse to within 5\%.

\begin{figure}[ht]
\label{fig1}
\centerline{\epsfxsize=3.375in\epsfbox{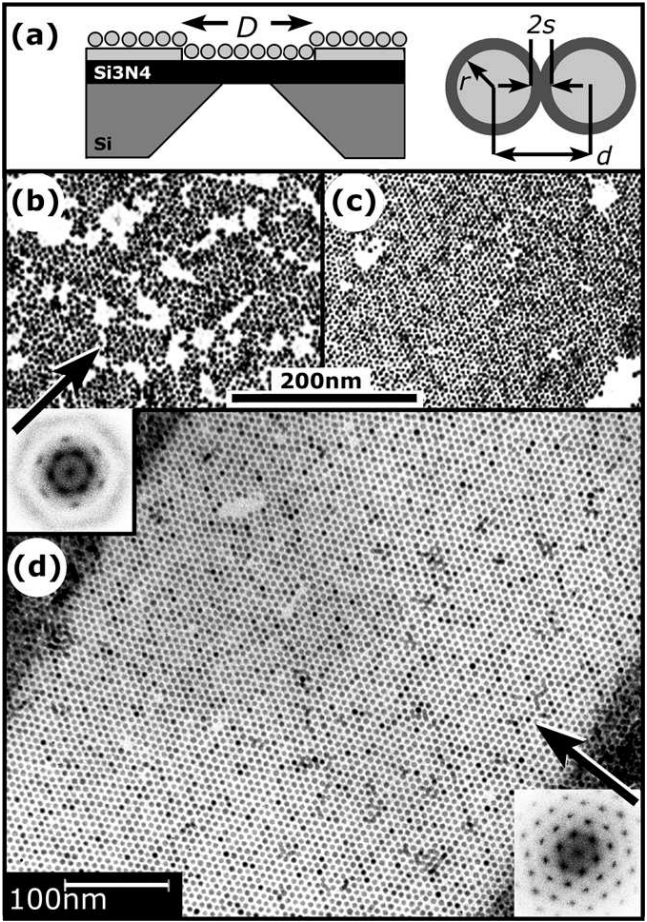}}
\caption{
(a) Sketch of a nanocrystal monolayer and
in-plane electrodes (not to scale), and of the interparticle
geometry.  (b)  TEM image (detail) of a typical array
formed {\em without} excess
dodecanethiol, showing many voids and an absence of long-range order.  The
2D Fourier Transform (2DFT) is of a larger area.  (c)  Typical
superlattice formed {\em with} excess dodecanethiol, 
showing $<$5\% voids and long-range ordering.
(d)  Highly ordered superlattice (and 2DFT) between
electrodes visible at the upper left and lower
right.  The random nanocrystal image intensities are due to
random Bragg diffraction of the electron beam.}
\end{figure}

Two different array preparation techniques yielded two distinct classes of
arrays:  samples with and without large-scale structural disorder.  Simple
deposition of 15--20 $\mu$l of colloid onto a substrate produced, 
upon drying,
a nanocrystal monolayer of well-packed regions with short-range order,
coexisting with numerous voids (area fraction 15--20\%) throughout the 2D
plane (Fig. 1b).  Higher particle concentrations produced an increased
amount of multi-layered regions without increasing the long-range
order.  However, addition of excess dodecanethiol (volume fraction
$6.3 \times 10^{-3}$) to the solution before deposition
increased the nanocrystal
mobility on the Si$_3$N$_4$ surface and 
prevented rapid dewetting of the solvent
from the substrate.  As a result, arrays with significantly smaller void
or double layer fraction (about 5\% combined) and excellent long-range
periodicity could be self-assembled  (Figs. 1c,d) \cite{lin2}.  Electronic
properties of a total of 14 arrays were measured (7 prepared without and 7
with excess ligand), with dimensions defined by the electrode separations
(200nm $< D <$ 700nm) and widths (fixed at $2 \mu$m).  
The resulting $N \times M$ arrays
ranged in length from $N =$ 30 to 90 particles and were 
$M \approx 270$ nanocrystals wide. 

The samples were cooled, in vacuum, to below 77K to avoid parasitic
conduction through the substrates.  DC current--voltage
($IV$) characteristics were measured using Keithley 614 electrometers and a
voltage source.  Control measurements on substrates without monolayers
showed no detectable currents ($<$ 0.02pA) up to $\pm20$V.   We imaged each
array by TEM after the transport measurements.  From analysis of the
center-to-center distances, $d$, and the particle radii, $r$, distributions of
the interparticle spacings $ 2s = d - 2r$ (Fig. 1a)  were obtained,
resulting in $s = 0.85 \pm 0.1$nm for the arrays with voids.  For the
well-ordered superlattices the excess dodecanethiol increased the mean
spacing to $s = 1.2 \pm 0.1$nm.

The $IV$ curves (Fig.2) were highly symmetric, of the same overall shape for
all samples, and showed no hysteresis at the slow ramp rates used ($<$5
mV/s).  The strongly non-ohmic behavior, in which current flow requires
the applied voltage to exceed a finite threshold, is characteristic of
Coulomb blockade of transport \cite{black,cordanetal,kurdak,rimbergetal,bez}.
The electrostatic energy
needed to add one electron to a quantum dot of charge $q$ and
(self-) capacitance $C_0 = 4\pi \epsilon \epsilon_0 r$ 
leads to a single-particle Coulomb
blockade voltage, $V_0 = q/C_0$, 
below which tunneling is suppressed (Fig. 2,
lower inset).  For $V>V_0$, current flows with resistance
$R = {\rm d}V/{\rm d}I$; $R \gg h/e^2$
for our nanocrystals.  The overall $IV$ characteristic then arises from the
series-parallel combination of many tunneling paths from particle to
particle throughout the array.  Random, parasitic charges  in the
substrate ---  local charge disorder --- 
induce the effective charges $q \in (0,e)$ on the nanocrystals, 
randomly placing the local $V_0$ in the interval
$(0,e/C_0)$, leaving $R$ unchanged.   Local structural disorder, on the other
hand, produces variations in the tunnel distance, $2s$, and thus a wide
(most likely log-normal) distribution of $R$, leaving $V_0$ essentially
unchanged \cite{snote}.  
The exponential dependence of $R$ on $s$ makes a large
variance in the interparticle resistances almost a certainty, even in
well-assembled arrays (e.g. Fig. 1d). 

Using the dielectric constant $\epsilon \approx 2$ 
for dodecanethiol, and $q = e/2$,  we find
$C_0 \approx 0.5$aF and $V_0 \approx 150$ mV for a 
typical Au nanocrystal.  The capacitance
between neighboring particles \cite{smythe}, 
based on the geometry observed by TEM,
is $C_{12} = 0.25$aF $ < C_0$.  
In this regime of small interparticle capacitive
coupling, hysteresis due to long-range charge re-arrangements is not
expected \cite{mw}.  Furthermore, for all accessible temperatures 
$e^2/\max\{C_0,C_{12}\} \gg {\rm k}_{\rm B}T$
and thus the array transport properties are essentially
temperature-independent (Fig. 2).  This behavior implicates direct,
interparticle quantum tunneling as the conduction mechanism and is in
contrast to the strong temperature dependence observed  in other
nanocrystal systems with large Coulomb blockade voltages \cite{black,andres}.
Arrhenius
behavior in those systems we believe is due to activated hopping via
trapping sites provided by electron-rich $\pi$-bonded linker molecules
attached to the metal particle cores.  This difference between tunneling
and activated hopping suggests that chemical modifications to
nanocrystalline systems may change fundamentally the physics governing
transport, and highlights two different routes towards control of
nanostructure conductance: physical tailoring of separations between metal
cores and chemical manipulation of molecular links between the cores.

\begin{figure}[tb]
\label{fig2}
\centerline{\epsfxsize=2.7in\epsfbox{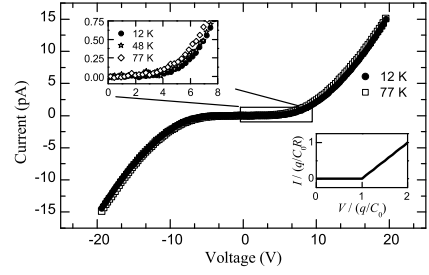}}
\floatsep=0.1in
\textfloatsep=0.1in
\dbltextfloatsep=0.1in
\intextsep=0.1in
\caption{
$IV$ curves for a typical superlattice ($D$ = 330nm $\times$ 
width $2\mu$m).  The upper inset magnifies the data.  The lower inset shows
a schematic $IV$ curve for a single nanocrystal, as described in the text.}
\end{figure}

How charge disorder alone affects the global $IV$ characteristics measured
across large arrays has been calculated by Middleton and Wingreen
(MW) \cite{mw}. They find a well-defined, global threshold voltage, 
$V_{\rm T}$, which
delineates a second order phase transition: for $V<V_{\rm T}$ 
the array current is
zero, and for $V>V_{\rm T}$ 
the array conducts with a power-law scaling form, 
$I \sim M(V_0/R) [(V-V_{\rm T})/V_{\rm T}]^\zeta$.  
In Figure 3 we plot the normalized current, $i = IR/(MV_0)$, 
as a function of normalized voltage above threshold, 
$ v = (V-V_{\rm T})/V_{\rm T}$, 
for all 14 measured arrays.  The array width $M$ was obtained
from the TEM images.  $R$ we adjusted to obtain collapse of the
curves.  Typical $R$ were around 50 and 300T$\Omega$ 
for arrays without and with
excess dodecanethiol, respectively, the increase corresponding to the
increase in average particle spacing described above.   These $R$ are in
accord with conducting-tip atomic force microscope studies of tunneling
through self-assembled alkanethiol monolayers \cite{wold}.  
Consistent choice of
$V_{\rm T}$ (not far from the 
values estimated by eye from the $IV$ curves,
e.g. Fig. 2) produced a robust scaling behavior of the {\em IV}s 
for the ordered
arrays assembled with excess ligands (Fig. 3a), yielding an average
$\zeta = 2.25 \pm 0.1.$

The conductivity exponent $\zeta$ is 
related to the meandering of current paths
in the charge-disordered landscape.  Even in regular 2D arrays current paths
are not straight, but exhibit transverse fluctuations that extend a
distance $\xi_\perp \propto v^{-\eta}$ due to the quenched disorder \cite{mw}.  
The total current above
threshold across an array of width $M$ is proportional to the number of
independent parallel paths, $M/\xi_\perp$, which leads to 
$i \propto vM/\xi_\perp \propto Mv^{\eta+1}$, i.e. $\zeta = \eta+1$. 
The determination of the transverse correlation length exponent $\eta$
arises in many physical phenomena connected to interface growth or
directed percolation.  MW argue that $\eta = 1/z$, where $z=3/2$ 
is the roughness
exponent for Kardar-Parisi-Zhang (KPZ) models for interface growth in 2D
\cite{kpz}; thus $\zeta \approx 5/3$ \cite{kpznote}.  
Simulations show $\zeta <\approx 2$ (i.e. $\eta <\approx 1$) for
square arrays of up to $400 \times 400$ 
particles \cite{mw}.  The power-law form of the
data in Fig. 3a suggests that scaling models are indeed
valid, but the measured
$\zeta$ 
is significantly larger than predicted by available theory.

\widetext
\begin{figure*}[t]
\centerline{\epsfxsize=6.5in\epsfbox{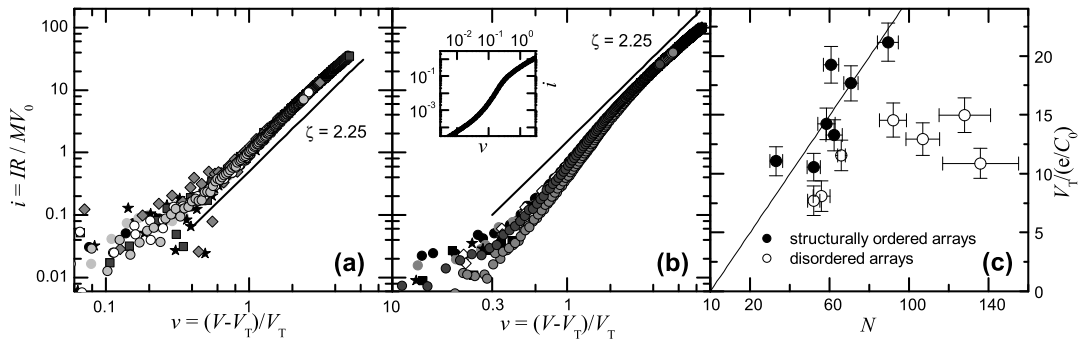}}
\label{fig3}
\floatsep=0.1in
\textfloatsep=0.1in
\dbltextfloatsep=0.1in
\intextsep=0.1in
\caption{
Scaling behavior of $IV$ curves.  (a,b) Log-log plots of all data
with normalized voltage beyond threshold $v = (V-V_{\rm T})/V_{\rm T} $
and current $i = IR/(MV_0)$ as described in the text.  
(a)  Data from seven monolayers with
long-range structural order.    The solid line shows best-fit power-law
$i=v^\zeta$ with $\zeta = 2.25$.  
(b)  Data from seven disordered
monolayers.  The solid line shows $\zeta = 2.25$ as in (a).
Inset: $IV$
curve from a (1+1)D simulation of 100 independent parallel channels each
100 particles in length.  
(c)  Threshold voltage $V_{\rm T}$, in units of $e/C_0$,
versus array length $N$.  The line  is a fit to 
$V_{\rm T} = \alpha N(e/C_0)$, which gives
$\alpha=0.25$ for the superlattices formed with excess dodecanethiol.}
\end{figure*}
\narrowtext

Several previous experiments have reported a wide, sample-dependent
spread in the scaling exponent:
$1.4< \zeta <2.0$
for lithographically patterned junction arrays \cite{kurdak,rimbergetal}, 
$1.6< \zeta < 2.1$ for polydisperse
nanocrystals \cite{wybourne}, and $2.2< \zeta < 2.7$ for 
small multi-layered arrays \cite{black}.  This contrasts with the
highly reproducible values we observe for $\zeta$ in the
structurally ordered arrays.  However, we find that
arrays with large void fraction display  $i(v)$ scaling behavior with
characteristic slope changes as well as larger
sample-to-sample variations (Fig. 3b).
The differences between the two classes of IVs may be 
understood in terms of
changes in network topology.
At large void fraction, neighboring voids produce
bottlenecks, locally cutting off  the
transverse correlation length $\xi_\perp$. 
In the extreme case, conduction is
reduced to several parallel 1D channels, each with linear 
$i \propto v$
but threshold $V_{\rm T}$ 
distributed over some range.  For $V$ barely larger than
the smallest $V_{\rm T}$ 
only one channel is open and the overall $IV$ is linear.  As
the applied voltage is increased, there will be a cross-over region in
which a growing number of parallel channels conduct.  This is
born out by the simulated $IV$ characteristic of such a (1+1)D system
(Fig. 3b, inset).  Depending on the number of independent channels,
behavior resembling power-laws in the cross-over region with exponents
ranging from $\zeta \approx 1$ to $\zeta > 2.5$ can easily be reproduced
\cite{1Dnote}.  
For sufficiently large $v$ all 1D channels have opened
and the overall $IV$ necessarily becomes linear again.   Actual arrays
most likely are amalgams of locally 2D patches connected by 1D channels.
Therefore, once all 1D bottlenecks are filled, the overall $IV$
characteristics are dominated by the remaining 2D patches and 
$\zeta$ approaches a value close to that of
ordered 2D arrays.   This is the behavior seen in Fig. 3b, where 
$\zeta \approx 2.7$ in the cross-over region  turns over to 
$\zeta = 2.16 \pm 0.1$ beyond $v \approx 2$.

Even for our largest accessible $v$, neither the {\em IV}s of the
ordered nor disordered arrays turn linear (Fig. 3a,b), as would be
expected once $\xi_\perp$ approaches a single lattice
spacing.  This finding is consistent with results from 
lithographically patterned arrays below $v \approx 10$ 
\cite{kurdak} and shows a remarkable extent of the scaling regime.

The thresholds $V_{\rm T}$ obtained from the 
scaling collapse of the ordered arrays
grow linearly with array length, $N$ (Fig. 3c):  
$V_{\rm T}/(e/C_0) = \alpha N$ with $\alpha = 0.25 \pm 0.02$.  
Simulations \cite{mw} show that $\alpha$ for a given 
lattice depends only on the
capacitive coupling between neighboring particles and decreases as
coupling increases; 2D square arrays in the limit 
$C_{12}/C_0 \rightarrow 0$ give $\alpha = 0.338$.  
Due to both the larger coordination number in our hexagonal
arrays and the finite $C_{12}/C_0 \approx 0.4$ 
we expect $\alpha < 0.338$, consistent with
the measured value.  For arrays with large void fraction, $N$ is poorly
defined, leading to strongly sample-dependent $V_{\rm T}(N)$ values.

From our findings, two key results emerge.
First, sufficiently large amounts of
topological disorder, due to voids in the monolayer, lead to distinct
deviations from simple power law behavior in the $IV$ characteristics.  In
situations where direct imaging is impossible, detailed examination of
{\em IV}s, therefore, may provide clues about an array's large scale
topology.  Second, $IV$ characteristics of long-range ordered arrays are
well-fit by a single power law, despite the existence of both charge
disorder and an inherent wide distribution of tunnel resistances.  We
believe this indicates that while $R$ may be exponentially sensitive to
variations in the interparticle separation, charge disorder nevertheless
plays the dominant role in selecting optimal current paths across the
array.  The reason most likely lies in the extremely non-linear, local $IV$
characteristics (Fig. 2, inset), which effectively shut off all current
flow unless $V_0$ is exceeded \cite{qnote}.   
Within this picture, our finding of an
exponent $\zeta \approx 2.25$ 
in the structurally well-ordered arrays implies $\eta \approx 1.15$ and
thus, as $V_{\rm T}$ is approached from above, a 
stronger divergence of $\xi_\perp$ than
would be expected from charge disorder alone (where $\eta \approx 0.67$).
In other words, the spread in tunnel 
resistances appears to produce a more rapid
growth of transverse fluctuations in the meandering current paths.

We thank  P. Guyot-Sionnest, L. P.  Kadanoff, and T. F.  Rosenbaum
for helpful discussions, and N. W. Mueggenburg and A. W. Smith for
experimental assistance.  This work was supported by the Keck Foundation
through Grant Number 991705, and by the MRSEC program of the National
Science Foundation under Award Number DMR-9808595.  One of us
(RP) acknowledges an NSF Graduate Fellowship, and a Grainger Graduate
Fellowship.


\begin {references}

\bibitem {timp} G. L. Timp, ed., {\em Nanotechnology}
(Springer-Verlag, New York, 1999).

\bibitem {collier97} C. P. Collier, R. J. Saykally, 
J. J. Shiang, S. E. Henrichs, and
J. R. Heath, Science {\bf 277}, 1978 (1997).

\bibitem {black} C. T. Black, C. B. Murray, R. L. Sandstrom, and 
S. Sun, Science {\bf 290}, 1131 (2000).

\bibitem {mw} A. A. Middleton and N. S. Wingreen, Phys. Rev. Lett. 
{\bf 71}, 3198 (1993).

\bibitem {collier98} C. P. Collier, T. Vossmeyer, and J. R. Heath,
Annu. Rev. Phys. Chem. {\bf 49}, 371 (1998).

\bibitem {andres} R. P. Andres {\em et al.}, Science {\bf273}, 1690 (1996)

\bibitem {cordanetal} A. S. Cordan {\em et al.}, J. Appl. Phys. {\bf 87}, 
345 (2000); C. Vieu {\em et al.}, J. Vac. Sci. Technol. B {\bf 16}, 3789 
(1998); W. Chen {\em et al.}, Appl. Phys. Lett. {\bf 66}, 3383 (1995).

\bibitem {wybourne} M. N. Wybourne {\em et al.}, Jpn. J. Appl. Phys. 
{\bf 36}, 7796 (1997).

\bibitem {morkved} T. L. Morkved {\em et al.}, Polymer {\bf 39}, 
3871 (1998).

\bibitem {lin1} X. M. Lin {\em et al.} J. Nanoparticle Res. {\bf 2}, 
157 (2000).

\bibitem {lin2} X. M. Lin, H. M. Jaeger, C. M. Sorensen, and 
K. J. Klabunde (to be published).  
Without electrodes present, superlattices routinely exhibit
long-range order over microns.  Electrodes perturb the monolayer
formation, leading to variability in superlattice quality.

\bibitem {kurdak} \c{C}. Kurdak {\em et al.}, 
Phys. Rev. B {\bf 57}, R6842 (1998).

\bibitem {rimbergetal} A. J. Rimberg {\em et al.}, Phys. Rev. Lett. 
{\bf 74}, 4714 (1995);  C. I. Duru\"{o}z
{\em et al.},  Phys. Rev. Lett. {\bf 74}, 3237 (1995).

\bibitem {bez} A. Bezryadin {\em et al.}, Appl. Phys. Lett. 
{\bf 74}, 2699 (1999).

\bibitem {snote} In the limit $C_{12}<C_0$ the effects of changes in $s$ on 
$C_{12}$ and $V_0$ \cite{mw} are
small compared to the exponential impact on $R$.

\bibitem {smythe} W. R. Smythe, {\em Static and Dynamic Electricity}
(McGraw-Hill, New York, 1950).

\bibitem {wold} D. J. Wold and C. D. Frisbie, J. Am. Chem. Soc. 
{\bf 122}, 1970 (2000).

\bibitem {kpz}  M. Kardar, G. Parisi, and Y.-C. Zhang, 
Phys. Rev. Lett. {\bf 56}, 889
(1986); S. Roux {\em et al.}, J. Phys. A {\bf 24}, L295 (1991).  
H. Hinrichsen,
Braz. J. Phys. {\bf 30}, 69 (2000), also {\em cond-mat}/9910284.

\bibitem {kpznote} 
The KPZ interface roughness exponent $z$ is the ratio of exponents
for fluctuations parallel ($\nu_\parallel$) and 
perpendicular ($\nu_\perp$) to the growth
direction.  Intriguingly, linking $\eta$ to transverse fluctuations only,
i.e. $\eta = \nu_\perp$ rather than $\eta = 1/z$,  and using
$\nu_\perp \approx 1$ \cite{kpz} gives $\zeta \approx 2$.

\bibitem {1Dnote} 
For example, for 100 parallel channels of 100 particles each, 
$\zeta = 2.7$ at the center of the crossover region.

\bibitem {qnote} Current can only flow through paths for which 
$V > \Sigma q_i/C_0$,
irrespective of whether such paths minimize the overall resistance across
the array.
  
\end {references}
\end{document}